\documentclass[lettersize,journal]{IEEEtran}
\usepackage[]{amsmath,amsfonts}
\usepackage{algorithmic}
\usepackage{array}
\usepackage[caption=false,font=normalsize,labelfont=sf,textfont=sf]{subfig}
\usepackage{textcomp}
\usepackage{stfloats}
\usepackage{url}
\usepackage{verbatim}
\usepackage{graphicx}
\usepackage{cite}
\usepackage{multirow}
\usepackage{siunitx}
\def\BibTeX{{\rm B\kern-.05em{\sc i\kern-.025em b}\kern-.08em
    T\kern-.1667em\lower.7ex\hbox{E}\kern-.125emX}}
\usepackage{balance}
\usepackage[inkscapeformat=png]{svg}

\usepackage{mathtools}
\usepackage{lipsum}
\usepackage{amsthm}
\usepackage{flushend}
\DeclareMathOperator*{\argmin}{argmin}
\theoremstyle{definition}
\newtheorem*{Proposition*}{Proposition}
\newtheorem{Assumption}{Assumption}
\newtheorem*{condition1*}{Convergence Consistency}
\newtheorem*{condition2*}{Solutions Consistency}

\makeatletter
\let\old@ps@headings\ps@headings
\let\old@ps@IEEEtitlepagestyle\ps@IEEEtitlepagestyle
\def\psccfooter#1{%
    \def\ps@headings{%
        \old@ps@headings%
        \def\@oddfoot{\strut\hfill#1\hfill\strut}%
        \def\@evenfoot{\strut\hfill#1\hfill\strut}%
    }%
    \def\ps@IEEEtitlepagestyle{%
        \old@ps@IEEEtitlepagestyle%
        \def\@oddfoot{\strut\hfill#1\hfill\strut}%
        \def\@evenfoot{\strut\hfill#1\hfill\strut}%
    }%
    \ps@headings%
}
\makeatother

\begin{document}

\title{On the Detection of Shared Data Manipulation\\ in Distributed Optimization}

\author{Mohannad Alkhraijah,
Rachel Harris,
Samuel Litchfield,
David Huggins,
and Daniel K. Molzahn\thanks{\noindent Support from NSF AI Institute for Advances in Optimization, \#2112533. Mohannad Alkhraijah is with the Computational Science Center, National Renewable Energy Laboratory, Golden, CO 80401, USA (email: mohannad.alkhraijah@nrel.gov). Rachel Harris, and Daniel K. Molzahn are with the School of Electrical and Computer Engineering, Georgia Institute of Technology, Atlanta, GA 30332 USA (e-mail: mohannad@gatech.edu; rharris94@gatech.edu; molzahn@gatech.edu). Samuel Litchfield and David Huggins are with the Cybersecurity, Information Protection, and Hardware Evaluation Research Laboratory, Georgia Tech Research Institute, Atlanta, GA 30332 USA (e-mail: samuel.litchfield@gtri.gatech.edu; david.huggins@gtri.gatech.edu)}}

\maketitle

\begin{abstract}
This paper investigates the vulnerability of the Alternating Direction Method of Multipliers (ADMM) algorithm to shared data manipulation, with a focus on solving optimal power flow (OPF) problems. Deliberate data manipulation may cause the ADMM algorithm to converge to suboptimal solutions. We derive a sufficient condition for detecting data manipulation based on the theoretical convergence trajectory of the ADMM algorithm. We evaluate the performance of the detection condition on three data manipulation strategies with various complexity and stealth. The simplest attack sends the target values and each iteration, the second attack uses a feedback loop to find the next target values, and the last attack uses a bilevel optimization to find the target values. We then extend the three data manipulation strategies to avoid detection by the detection conditions and a neural network (NN) detection model. We also propose an adversarial NN training framework to detect shared data manipulation. We illustrate the performance of our data manipulation strategy and detection framework on OPF problems. The results show that the proposed detection condition successfully detect most of the data manipulation attacks. However, the bilevel optimization attack strategy that incorporates the detection methods may avoid being detected. Countering this, our proposed adversarial training framework detects all the instances of the bilevel optimization attack. 
\end{abstract}


\section{Introduction} \label{sec1:introduction}
Distributed optimization algorithms allow multiple agents to collaboratively solve large-scale optimization problems. Agents using distributed optimization solve subproblems iteratively and exchange the solutions of the shared variables with their neighbors at each iteration. If the solutions are correct and accurate, the algorithm converges to the optimal solution under mild technical assumptions. Using distributed algorithms to solve power system optimization problems, such as optimal power flow (OPF), has the potential to scale computations, increase reliability, and improve data privacy~\cite{molzahn2017survey}. 

Since distributed algorithms relay on communicating shared variables, these algorithms are vulnerable to communication nonidealities and data manipulation. Most of the existing literature on distributed optimization assumes that the participating agents are trustworthy and the shared data are accurate. Nonetheless, communications are prone to errors, and agents may deliberately share inaccurate solutions, e.g., to increase their profit by changing their local generators' outputs.

\subsection{Related Work}

Communication nonidealities significantly impact the performance of distributed algorithms. Even with a low probability of occurrence, large errors can prevent the algorithm from converging~\cite{ALKHRAIJAH2022108297}. Malicious agents may inject random errors to execute ``denial-of-service'' attacks that cause the algorithm to diverge. However, making the algorithm converge to a suboptimal solution while being stealthy is more challenging and requires deliberate manipulations.

Deliberate data manipulations of distributed optimization are plausible, but limited research investigates this type of attacks. Reference~\cite{7762902} considers false data injection attacks on a distributed algorithm that solves OPF problems with the DC power flow approximation (DCOPF). Reference~\cite{7762902} also proposes a detection method that estimates the shared variables using data communicated in previous iterations. The agents then update a reputation index for the neighboring agents based on the deviation from the expected values of the shared data. An extension in~\cite{9539887} considers OPF problems for radial networks using a second-order cone programming relaxation. Other work in~\cite{9762779} proposes the use of power line communication and encryption to prevent ``man-in-the-middle'' attacks that compromise the communication between the agents. However, the method in~\cite{9762779} only detects attacks on the communication between agents but not on the agent's controllers. Moreover, the methods in~\cite{7762902, 9539887,9762779} consider a simplistic attack scenario in which the attacker repeatedly shares data corresponding to a malicious target solution.

Our prior work introduced and analyzed various data manipulation strategies that drive distributed optimization algorithms to suboptimal solutions~\cite{tpec2022, harris_molzahn-hicss2024, 10012244}. We consider a stealthy attack model in~\cite{10012244} that uses a feedback loop to reduce the deviation of the manipulated data from the expected value. We also propose an attack model that uses bilevel optimization with the other subproblems in the lower level. The bilevel optimization attack model finds a solution that optimizes the attacker's malicious objective and ensures the convergence of the algorithm in two iterations. Due to the fast convergence of this attack, reputation-based detection methods may fail since these methods require multiple iterations before detecting an attack. This paper extends our previous work by considering a new detection method and more sophisticated attack strategies.

Using neural network (NN) models is a promising approach for detecting data manipulation~\cite{anomaly_survey}. Our prior work in~\cite{10012244, harris_molzahn-hicss2024} shows that NN detection models have the potential to detect a variety of data manipulation strategies in the context of distributed optimization algorithms. However, using NN models for anomaly detection is challenging due to the nature of anomalies, as they are rare and heterogeneous, and thus difficult to collect on a large scale for training~\cite{anomaly_survey}.

The detection models discussed above are data-driven with no detectability guarantees. Conversely,~\cite{munsing2018cybersecurity} proposes an analytical detection method for convex problems. The method in~\cite{munsing2018cybersecurity} estimates the Hessian matrix of the subproblems' augmented Lagrangian using the shared data. If the estimated Hessians are not positive semidefinite, which is necessary for convexity, then there is data manipulation. This method requires shared data from a large number of iterations. Moreover, numerical issues may arise due to poor matrix conditioning when estimating Hessian matrices, especially for solutions that are close to the optimal solution. Thus, the method requires shared data from a large number of iterations to improve conditioning at the expense of more computations. Nevertheless, these challenges may cause inaccurate estimates, potentially resulting in false positives. Moreover, the detection method in~\cite{munsing2018cybersecurity} is limited to convex problems.

\subsection{Contributions}

This paper investigates the vulnerability of the Alternating Direction Method of Multipliers (ADMM) algorithm to shared data manipulation when solving OPF problems. This paper proposes and evaluates several data manipulation and detection strategies. The main contributions of the paper are:

\noindent (1) Development of an analytical detection condition based on the convergence trajectory of the ADMM algorithm to check the correctness of the shared data. This condition verifies that the shared data corresponds to a solution for an optimization problem consistent with the solutions from previous iterations. Unlike existing data-driven approaches, we analytically derive this detection condition. The detection condition uses the shared variables of a single agent from any three consecutive iterations. Thus, the other agents can use the condition during any intermediate iteration to identify the malicious agents. Moreover, the detection condition applies to convex and non-convex problems with non-differentiable objectives. To the best of our knowledge, this is the first detection condition with these properties.

\noindent (2) Proposal of sophisticated data manipulation attacks that use bilevel optimization to bypass detection methods. Unlike the existing simple threat models in the literature, the proposed attack incorporates an NN detection model and the analytical detection condition that we propose in this paper. These attacks embed the detection methods into the attacker's bilevel optimization problem using mixed-integer linear programming (MILP) constraints.

\noindent (3) Proposal of an adversarial NN training framework to improve the detectability of data manipulation. The adversarially trained NN detects the proposed data manipulation attacks and can be extended to detect new attacks by retraining the NN model to enhance detection accuracy. We show that even when the attacker has access to the detection methods, bypassing the detection methods is computationally expensive, and thus renders the attacks ineffective. We also show that the analytical detection condition increases the efficiency of NN training by reducing the number of feasible attacks.

\subsection{Organization}

This paper is organized as follows. Section~\ref{sec2:background} presents the OPF problem and the ADMM algorithm. Section~\ref{sec3:detection_condition} proposes an analytical detection condition. Section~\ref{sec4:embed_detection} describes three data manipulation attacks that bypass detection methods. Section~\ref{sec5:adv_nn} presents an NN training framework to detect data manipulation. Section~\ref{sec6:results} presents numerical results of the proposed methods. Section~\ref{sec7:conclusion} gives conclusions.

\section{Background} \label{sec2:background}

This section presents the background information and notation for the OPF problem and the ADMM algorithm.

\subsection{Optimal Power Flow}

OPF is a fundamental optimization problem in power systems that finds the controller setpoints that minimize operational cost subject to the power flow equations and engineering constraints. There is a wide variety of OPF formulations, including various approximations and relaxations~\cite{molzahn_hiskens-fnt2019}. We present a general OPF formulation with the DC approximation for illustrative purposes, but the results in this paper apply to other OPF formulations. The DCOPF problem is
\begin{subequations}%
\label{eq:dcopf}%
\begin{align}%
& \mbox{min} \quad \sum_{g\in \mathcal{G}} c_{g_2} (p^G_{g})^2 + c_{g_1} p^G_{g} + c_{g_0} \label{eq:objective} &\\
& \mbox{subject to: }  \nonumber & \\
&\sum_{\substack{g \in \mathcal{G}_i}} p^G_{g} - \sum_{\substack{l \in \mathcal{L}_i}} p^L_{l} = \sum_{\substack{(i,j)\in \mathcal{E}}}  p^E_{ij}, &&\forall i\in \mathcal{N}, \label{eq:power_balance}  \\ 
& p^E_{ij} = b_{ij} (\theta_{i} - \theta_{j}), &&\forall (i,j)\in \mathcal{E}, \label{eq:power_flow} \\
& p_{g}^{min} \leq p^G_{g} \leq p_{g}^{max}, &&\forall g \in \mathcal{G}, \label{eq:generator_bounds} \\ 
&|p^E_{ij}| \leq p^{max}_{ij}, &&\forall (i,j) \in \mathcal{E}, \label{eq:branch_bounds} \\ 
& \theta_{r} = 0,   \label{eq:references}
\end{align}
\end{subequations}
\noindent where $\mathcal{N}$, $\mathcal{E}$, $\mathcal{G}$, and $\mathcal{L}$ are the sets of buses, branches, generators, and loads, respectively. The subsets $\mathcal{G}_i\subset\mathcal{G}$ and $\mathcal{L}_i\subset\mathcal{L}$ are the generators and loads connected to bus $i\in \mathcal{N}$. The decision variables are the buses' voltage phase angles $\theta \in\mathbb{R}^{|\mathcal{N}|}$, the generators' active power outputs $p^G\in\mathbb{R}^{|\mathcal{G}|}$, and the branches' active power flows $S\in\mathbb{R}^{|\mathcal{E}|}$, where $|\,\cdot\,|$ denotes the cardinality of a set. We denote the load demands with $p^L\in\mathbb{R}^{|\mathcal{L}|}$. We use $b_{ij}\in\mathbb{R}$ to denote the susceptance of branch $(i,j)\in\mathcal{E}$. Generator $g \in \mathcal{G}$ has a quadratic cost function with coefficients $c_{g_2}$, $c_{g_1}$, and $c_{g_0}$.

The objective~\eqref{eq:objective} minimizes the generation cost. The equality~\eqref{eq:power_balance} enforces power balance, and~\eqref{eq:power_flow} defines the branches' power flow. Inequality~\eqref{eq:generator_bounds} bounds the generators' power output between $p_g^{min}$ and $p_g^{max}$, and \eqref{eq:branch_bounds} bounds the branches power flow below $p_{ij}^{max}$. Equality~\eqref{eq:references} sets the reference angle in the chosen reference bus $r$.

For notational simplicity, we group the variables in a vector $x=[\theta^\mathsf{T}~(p^{G})^\mathsf{T}~(p^E)^\mathsf{T}]^\mathsf{T}$, where $(\,\cdot\,)^\mathsf{T}$ is the transpose operator. We denote the inequality and equality constraints in~\eqref{eq:power_balance}--\eqref{eq:references} as $h^E(x) = 0$ and $h^I(x) \leq 0$. We further define the set of feasible solutions as $\Omega = \{x | h^E(x) = 0,\, h^I(x) \leq 0 \}$.

\subsection{Alternating Direction Method of Multipliers}
ADMM is a well-known distributed optimization algorithm based on the augmented Lagrangian method~\cite{boyd2011distributed}. We first present the general form of the ADMM algorithm and then describe a special case that solves the consensus problem.

\subsubsection{General Form}
The ADMM algorithm solves problems in the form:
\begin{subequations}%
\label{eq:form}%
\begin{align}%
   & \min_{x, z} \qquad\qquad f(x) + g(z) \\
   & \mbox{subject to:} \quad\;\; Ax + Bz = c, \label{eq:consistency} \\
   & \phantom{subject to:} \quad\; x\in \mathcal{X},\; z\in \mathcal{Z},
\end{align}%
\end{subequations}%
\noindent where $x\in\mathbb{R}^n$ and $z\in\mathbb{R}^m$ are decision variables constrained to the nonempty, closed, convex sets~$\mathcal{X}$ and $\mathcal{Z}$, and $A\in\mathbb{R}^{p\times n}$, $B\in\mathbb{R}^{p\times m}$, and $c\in \mathbb{R}^p$ are the consistency constraint parameters. The functions $f\colon\mathcal{X}\rightarrow \mathbb{R}$ and $g\colon\mathcal{Z}\rightarrow \mathbb{R}$ are proper convex functions. The formulation~\eqref{eq:form} depict problems with two sets of decision variables $x$ and $z$ that have separable objective functions and linear coupling constraints.

The ADMM algorithm uses an augmented Lagrangian function to relax the consistency constraints~\eqref{eq:consistency}:
\begin{equation}%
L_{\rho}(x,z,y) = f (x) + g(z) + y^\mathsf{T} (Ax + Bz - c) + \frac{\rho}{2} ||Ax + Bz - c||^2_2,\nonumber
\end{equation}%
\noindent where $y\in \mathbb{R}^p$ are dual variables, $\rho\in \mathbb{R}_{>0}$ is tuning parameter, and $||\,\cdot\,||_2$ denotes the $l_2$-norm. The ADMM algorithm solves the augmented Lagrangian problem by alternatively solving for $x$ and $z$, and then updating the dual variables using a gradient ascending step. Thus, iterate $k+1$ solutions are:
\begin{subequations}%
\label{eq:admm}%
\begin{align}%
    &\!\! \begin{aligned} x^{k+1} \!\! \coloneq \argmin_{x\in\mathcal{X}}  f(x) \! + \! (y^k)^\mathsf{T} Ax+\frac{\rho}{2} ||Ax + Bz^k \! - \! c||^2_2, \end{aligned}\label{eq:admm1} \\  
    &\!\! \begin{aligned} z^{k+1} \!\! \coloneq \argmin_{z\in\mathcal{Z}} g(z) \! + \! (y^k)^\mathsf{T} Bz \! + \! \frac{\rho}{2} ||Ax^{k+1} \! + \! Bz \! - \! c||^2_2, \end{aligned} \label{eq:admm2} \\ 
    & \! y^{k+1} \! \coloneq y^k + \rho (Ax^{k+1} + Bz^{k+1} - c). \label{eq:admm3}
\end{align}
\end{subequations}%
Defining the primal residual $r^k = Ax^k + Bz^k - c$ and the dual residual $s^k = \rho A^\mathsf{T} B (z^{k} - z^{k-1})$ at iteration $k$, the algorithm terminates when the norms (often $l_2$ or $l_\infty$) of the primal and dual residuals are below a predefined tolerance.

\subsubsection{Consensus Problem} \label{sec2.B.2:consensus_problem}

Consensus problems are a special case of~\eqref{eq:form} that have multiple subproblems with separable objective functions and global variables. Let $\mathcal{A}$ be the set of agents solving the subproblems. We denote the decision variables of agent~$i\in\mathcal{A}$ as $x_i\in\mathbb{R}^{n}$, constrained to a convex set $\mathcal{X}_i$. We introduce auxiliary variables $z\in \mathbb{R}^{n}$ and enforce consistency between the shared variable by equating the same local variables with an auxiliary variable. The consensus optimization problem is:
\begin{subequations}%
\label{eq:consensus_form}%
\begin{align}%
   & \min_{x,z} && \sum_{i\in\mathcal{A}} f_i(x_i) \\
   & \mbox{subject to: }  && x_i - z = 0, &  \qquad\qquad \forall i\in\mathcal{A},\\
   & \phantom{subject to: }  && x_i\in\mathcal{X}_i, &  \qquad \forall i\in\mathcal{A}.
\end{align}
\end{subequations}%
\noindent Problem~\eqref{eq:consensus_form} is a special case of~\eqref{eq:form} with $f(x) = \sum_{i\in\mathcal{A}} f_i(x_i)$, $g(z) = 0$, $c = 0$, $A = I$, where $I$ is the identity matrix of appropriate size, and $B$ consists of $|\mathcal{A}|$ vertically stacked identity matrices multiplied by $-1$. The ADMM algorithm then solves the consensus problem via solving the following:
\begin{subequations}%
\label{eq:consensus_admm}%
\begin{align}%
    &x_i^{k+1} \!\! \coloneq \! \argmin_{x_i\in\mathcal{X}_i} f_i(x_i)\! + \! (y_i^k)^\mathsf{T} \! x_i \! + \!\! \frac{\rho}{2} ||x_i \! - \! z^k||^2_2, \forall i \in \mathcal{A},  \label{eq:consensus_admm1}  \\  
    &z^{k+1} := \frac{1}{|\mathcal{A}|} \sum_{i\in\mathcal{A}} x_i^{k+1}, \label{eq:consensus_admm2} \\ 
    &y_i^{k+1} := y_i^k + \rho (x_i^{k+1} - z^{k+1}), \qquad\qquad\qquad\; \forall i \in \mathcal{A}, \label{eq:consensus_admm3}
\end{align}
\end{subequations}%
\noindent where $y_i\in\mathbb{R}^{n}$ are agent~$i$ dual variables. 

To solve OPF problems~\eqref{eq:dcopf} using the ADMM algorithm~\eqref{eq:consensus_admm}, consider the case where there are multiple agents defined by the set~$\mathcal{A}$, each of which operates a partitioned subset of the buses $\mathcal{N}_i \subset \mathcal{N}, i\in\mathcal{A}$ along with the corresponding connected elements, i.e., generators $\mathcal{G}_j$ and loads~$\mathcal{L}_j$, $\forall j\in \mathcal{N}_i$. Although the objective function of the OPF problem~\eqref{eq:objective} is separable, the variables and constraints are coupled. To eliminate the coupling, we introduce auxiliary variables corresponding to the power flows of each branch that connects two partitions and the voltages of the boundary buses. Each agent takes a copy of the coupled constraints and variables in addition to consistency constraints between the coupled variables and the corresponding auxiliary variables. The consensus OPF problem thus has the form of~\eqref{eq:consensus_form} with $\mathcal{X}_i = \Omega_i, \forall i\in\mathcal{A}$, i.e., the set of OPF constraints in~\eqref{eq:dcopf}, and can be solved using the ADMM algorithm~\eqref{eq:consensus_admm}. The ADMM convergence guarantee is limited to convex problems, such as the DCOPF problem in~\eqref{eq:dcopf}. Nonetheless, empirical results show that the ADMM algorithm frequently converges to good solutions for non-convex OPF problems~\cite{6917065}.

\section{Detection Condition\\for Anomalous Shared Data} \label{sec3:detection_condition}

This section presents a sufficient condition to detect inconsistencies in the solutions shared by the agents when using the ADMM algorithm. The detection condition uses the fact that the agents repeatedly solve the same subproblem with different parameters. We exploit the subproblem structure to derive a necessary condition for the solutions of the subproblems that only depends on the shared variables, which are not private. When an agent violates this condition, this agent must not be solving the same subproblem, thus yielding a sufficient condition to detect data manipulation. Before stating the detection condition, we have the following two assumptions: 
\begin{Assumption} 
    The objective function $f$ is lower semi-continuous, but not necessarily convex, over the constraint set $\mathcal{X}$~\cite[Def. 1.5]{rockafellar2009variational}.
\end{Assumption}
\begin{Assumption}
    The constraint set $\mathcal{X}$ satisfies the linear independence constraint qualification (LICQ) at $x^{k}$ for any $k\geq1$~\cite[Def. 12.4]{nocedal1999numerical}
\end{Assumption}
Using these two assumptions, we state the sufficient detection condition in the following proposition. 
\begin{Proposition*}[Sufficient Detection Condition]
    Let $x^k$ and $z^k$ be the iterate~$k$ solutions of the ADMM algorithm~\eqref{eq:admm} and the problem satisfies assumptions 1 and 2. Define $\hat{z}^k = 2 B z^k - B z^{k-1} - c$. If $(x^{k+1} + A^{-1} \hat{z}^k)^\mathsf{T} A^\mathsf{T} A (x^{k+1} - x^k) \geq \epsilon$ for any $k\geq 1$ and small $\epsilon\in\mathbb{R}_{>0}$, then there is data manipulation. 
\end{Proposition*}
We derive the detection condition using the local subproblems' first-order optimality conditions. Since the agents share optimal solutions at each iteration, we can validate solution optimality by cross evaluating their objective functions with the current and previous shared variables.  The complete proof of the sufficient detection condition is in the appendix. 

The value of $\epsilon$ depends on the optimality tolerance of the numerical solutions. To preclude false positives, i.e., avoid flagging an attack while there is no data manipulation, we must select $\epsilon$ to be higher than the solver's numerical tolerance. 

For the consensus problem~\eqref{eq:consensus_form}, we evaluate the consistency parameters $A$, $B$, and $c$ as described in Section~\ref{sec2.B.2:consensus_problem}. The condition thus becomes
\begin{equation}
(x_i^{k+1} - 2z^k + z^{k-1})^\mathsf{T} (x_i^{k+1} - x_i^k) \leq 0, \qquad \forall i\in \mathcal{A}. \label{eq:condition2}
\end{equation}

The two assumptions we used in the proposition are typical for NLP problems and generally true for OPF problems~\cite{8391733}. An implicit assumption we used is the ability of agents to solve the subproblem to global optimality. This assumption is hard to validate for non-convex problems. Nonetheless, empirical studies such as~\cite{gopinath2022} show that local NLP solvers often find good solutions with small optimality gaps for many OPF instances. If an upper bound on the optimality gap is known, we can incorporate this bound in the value of $\epsilon$ when solving NLP problems and the proposition holds. 

Thus, this condition is sufficient for detecting data manipulation in non-convex problems with non-differentiable objectives, only requires the values of shared variables from three consecutive iterations, and can be checked at each iteration. Moreover, since we can evaluate this condition for each agent separately, the condition identifies the agents that manipulate the shared data. To the best of our knowledge, this is the first sufficient condition with these advantages.

\section{Data Manipulation with\linebreak Embedded Detection Models} \label{sec4:embed_detection}
The detection condition in Section~\ref{sec3:detection_condition} detects various types of data manipulations, as Section~\ref{sec6:results} shows later in the paper. However, a sophisticated attacker with knowledge of the detection methods could incorporate the detection condition into their attack model to compute attacks that avoid detection. To analyze these sophisticated attacks, this section first reviews three data manipulation models from~\cite{10012244} that steer the ADMM algorithm to malicious operating points. Then, this section describes how an attacker could use the detection condition and an NN detection model to remain undetected.

\subsection{Data Manipulation Strategies}
We build on the attack model proposed in~\cite{10012244}, which presents three data manipulation strategies that drive the results of the distributed algorithm to a malicious operating point. Throughout this section, we denote the attacker as agent~$a$.

\subsubsection{Simple Attacker Model}
In this model, the attacker simply shares the target values of the shared variables at each iteration. This model requires finding a target point that is feasible for neighboring agents before the attack starts. The simple attack is easily detected because the attacker sends the same values in each iteration.

\subsubsection{Feedback Attacker Model}
The feedback attack model is a generalization of the simple attack model that is harder to detect. Instead of repeatedly sending the exact target values, the attacker uses a proportional–integral–derivative (PID) feedback loop to compute the shared variables. Similar to the simple attack, the attacker determines the desired target values of the shared variables in advance. Let $\hat{x}_a$ be the attacker's target values, $x_a$ be the shared variables that the attacker actually sends, and $z_a$ be the shared variables defined in~\eqref{eq:consensus_admm}. The attacker computes the shared variables at iteration $k+1$: 
\begin{equation} \label{eq:feedback}
x^{k+1}_{a} = \hat{x}_a + K_p e_k + K_i \sum_{n=1}^{k} e_n + K_d (e_k - e_{k-1}),
\end{equation}
\noindent where $K_p$, $K_i$, and $K_d$ are the PID feedback tuning parameters and $e_k = \hat{x}_a - z_a^k$ is the deviation of the neighboring agents' shared variables from the attacker's target values. Choosing $K_p = K_i = K_d = 0$ yields the simple attack model.

\subsubsection{Bilevel Attacker Model}
\begin{figure}[t]
\vspace{-10pt}
    \centering
        \includegraphics[width=2.6in]{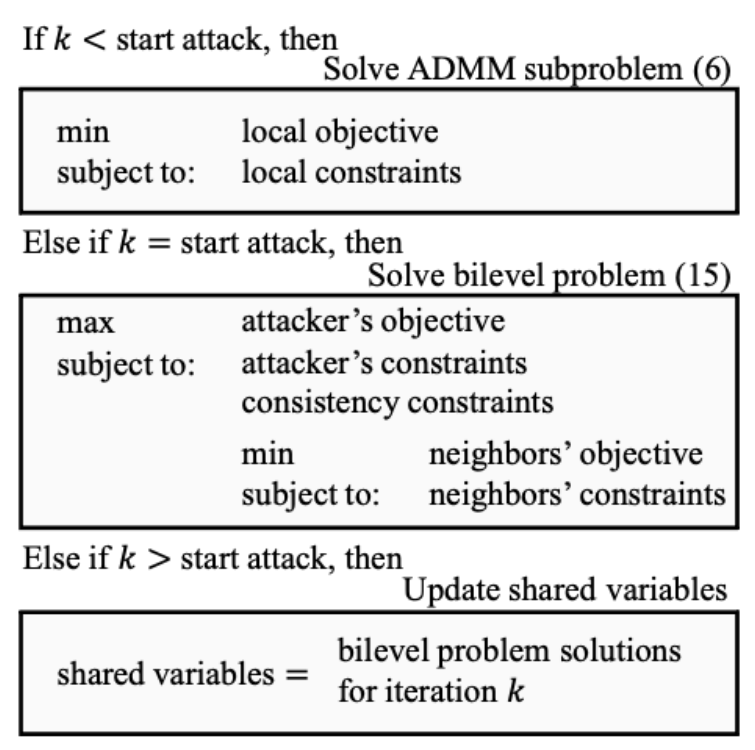}
	\caption[attacker]{The attacker's logic when using the bilevel model. The attacker selects a starting iteration. Before starting the attack, the attacker solves normal ADMM subproblems~\eqref{eq:consensus_admm}. After starting the attack, the attacker shares manipulated data obtained by solving the bilevel problem~\eqref{eq:bilevel}.}
	\label{fig:attacker_logic}
\end{figure}

This model involves solving a bilevel optimization that includes the subproblems of the neighboring agents in the lower level and the attacker's objective and constraints in the upper level. As Fig.~\ref{fig:attacker_logic} illustrates, the attacker computes the shared data at a selected start iteration~$s$ and the next two iterations by solving~\eqref{eq:bilevel}:
\begin{subequations}%
\label{eq:bilevel}%
\begin{align}%
   & \min_{x,y,z}  \qquad F(x_a^{s+2}) \label{eq:bilevel_objective}\\
   & \mbox{subject to:} \nonumber \\
   & x^k_i = \argmin_{x_i\in \Omega_i} f_i(x_i) + (y_i^{k-1})^\mathsf{T} x_i +  \frac{\rho}{2} ||x_i - z_i^{k-1}||, \nonumber \\ 
   & \mspace{161.8mu} \forall i \in \mathcal{A}_a, k \in \{s,s+1,s+2\}, \label{eq:lower_level}\\
   & z^{k} =  \frac{1}{|\mathcal{A}|} \! \sum_{i\in\mathcal{A}} x_i^{k}, \qquad\qquad\;\; \forall i\in \mathcal{A}_a,k\in\{s,s+1\},  \label{eq:bilevel_shared}\\
   & y_i^{k} = y_i^{k-1} + \rho (x_i^{k} - z_i^{k}), \mspace{17.5mu}\forall i \in \mathcal{A}_a, k\in\{s,s+1\}, \label{eq:bilevel_dual} \\
   & x_a^{s+2} \in \Omega_a, \label{eq:bilevel_feasiability} \\
   & x_a^{s+2} = x_i^{s+2},  \mspace{218.4mu} \forall i \in \mathcal{A}_a, \label{eq:bilevel_consistancy}
\end{align}
\end{subequations}%
\noindent where $F\colon\Omega_a \rightarrow \mathbb{R}$ is the attacker's objective function describing how they would like to manipulate the solution of the distributed optimization algorithm (e.g., increasing local generation in the attacker's area). The set $\Omega_i$ denotes the OPF constraints for agent~$i$ as defined in~\eqref{eq:dcopf}. The set $\mathcal{A}_a$ consists of the attacker's neighboring agents. 

The lower-level problems in~\eqref{eq:lower_level} are the local subproblems of the neighboring agents for three consecutive iterations. The first iteration $s$ can be calculated in advance since the values of $z^{s-1}$ and $y^{s-1}$ are known, but we keep this iteration in the formulation to illustrate the need to find the corresponding $x^s_i$, $\forall i\in \mathcal{A}_a$. Constraints~\eqref{eq:bilevel_shared}--\eqref{eq:bilevel_dual} are the ADMM updates for iterations $s$ and $s+1$. Constraints~\eqref{eq:bilevel_feasiability} and~\eqref{eq:bilevel_consistancy} ensure that the attacker's solution is feasible and the consensus is achieved at iteration $s+2$.

The complexity of the bilevel problem~\eqref{eq:bilevel} depends on the power flow model used in the lower-level subproblems~\eqref{eq:lower_level} and the form of the upper-level objective~\eqref{eq:bilevel_objective}. With a linear power flow approximation and a linear objective function $F$, both the lower- and upper-level problems are linear programs. We can then solve the bilevel problem via reformulating the lower-level using the Karush–Kuhn–Tucker (KKT) conditions parameterized with the upper-level variables $y$ and $z$. Then, we can use a standard big-M method~\cite{RepresentationAmet1981} or a \mbox{type-1} special ordered set (SOS1) method~\cite{vielma2011modeling} to reformulate the KKT conditions as MILP constraints.

\subsection{Embedding Detection Models}
This section extends the three data manipulation strategies we previously described to incorporate the SC detection condition in Section~\ref{sec3:detection_condition} and an NN detection model. We do not consider the CC condition described in Section~\ref{sec3:detection_condition} because of its dependence on the final solution and, as we will show in Section~\ref{sec6:results}, we found that the SC condition is more accurate.

\subsubsection{Embedding Detection Conditions}

To bypass the detection condition described in Section~\ref{sec3:detection_condition}, the attacker imposes additional constraints on their shared variables to ensure satisfying~\eqref{eq:condition2}. For the simple and feedback attack models, a sophisticated attacker projects the shared variables computed by the attacks onto the feasible region of the detection condition~\eqref{eq:condition2} by solving the following problem:
\begin{subequations}\label{eq:embd_feedback}
\begin{align}
& \min_{x_a}  \qquad ||x_a - \check{x}_a||^2_2 \label{eq:embd_feedback1}\\
& \mbox{subject to:} \nonumber \\ 
& (x_a  - 2z^k + z^{k-1})^\mathsf{T} (x_a - x_a^{k}) \leq 0, \label{eq:embd_feedback2} \\ 
& \check{x}_a = \hat{x}_a + K_p e_k + K_i \sum_{n=1}^{k} e_n + K_d (e_k - e_{k-1}), \label{eq:embd_feedback3}
\end{align}
\end{subequations} 
\noindent where $x_a$ is the vector of the attacker's shared variable, $z_a$ is the vector of shared variables as defined in~\eqref{eq:consensus_admm}, $\hat{x}_a$ is the vector of target shared variables, $\check{x}_a$ is the output of the feedback attack in~\eqref{eq:feedback}, and $e_n = \hat{x}_a^n - z_a^n$, $n\in \{1,2,\ldots, k\}$, are error terms. The objective function~\eqref{eq:embd_feedback1} finds the closest shared variables to the output of the feedback attack~$\check{x}_a$. Constraint~\eqref{eq:embd_feedback2} ensures that the detection condition will not detect the attack, and~\eqref{eq:embd_feedback3} is the output of the feedback attack. 

Letting $K_p = K_i = K_d = 0$, we obtain the simple attack model with embedded detection conditions. The simple attack in this case is a projection of the attacker's target values onto the closest value that cannot be detected by the detection condition. 

Due to~\eqref{eq:embd_feedback2}, problem~\eqref{eq:embd_feedback} is a non-convex quadratic problem that the attacker can solve using an NLP solver to find a local solution. Any feasible solution to~\eqref{eq:embd_feedback} that satisfies~\eqref{eq:condition2} will bypass the detection condition. Thus, locally optimal solutions are sufficient to drive the ADMM algorithm to the attacker's target values while avoiding detection. However, the solutions of~\eqref{eq:embd_feedback} are not necessarily the best possible attack.

For the bilevel attack model, the attacker adds constraints to~\eqref{eq:bilevel} that ensure the satisfaction of the detection condition~\eqref{eq:condition2}. The additional constraints are
\begin{equation}\label{eq:bilevel_additional}
    (x_a^{k}  - 2z^{k-1} + z^{k-2})^\mathsf{T} (x_a^{k} - x_a^{k-1}) \leq 0
\end{equation}
\noindent for iterations $k\in \{s,s+1,s+2\}$, where $x_a^{s}$, $x_a^{s+1}$, and $x_a^{s+2}$ are the attacker's shared variables; $z_a^{s}$ and $z_a^{s+1}$ are the shared variables as defined in~\eqref{eq:consensus_admm}; and $x_a^{s-1}$, $z_a^{s-1}$, and $z_a^{s-2}$ are the shared variables from the prior iterations. As a non-convex quadratic inequality, enforcing~\eqref{eq:bilevel_additional} in the upper level increases the attacker's computational challenges when solving the bilevel problem~\eqref{eq:bilevel}. Nonetheless, commercial solvers like Gurobi are applicable to this type of quadratically constrained problem when appropriately formulated.

\subsubsection{Embedding Neural Network Detection Model} \label{subsec:embedding_NN}
Our previous work in~\cite{10012244} presents an NN detection model that shows promising results in detecting attacks. To avoid detection by the NN model in~\cite{10012244}, this section presents a data manipulation attack that embeds the NN in the bilevel problem by adding constraints ensuring that the NN indicates no attack has occurred. For the simple and feedback attacks, embedding NN detection models into the attack strategies is more challenging, as we will discuss at the end of this section. 

Consider an NN with a set $\mathcal{U}$ of layers indexed from $1$ to $|\mathcal{U}|$, each with a set $\mathcal{V}_{u}, u \in \mathcal{U}$, of neurons indexed from $1$ to $|\mathcal{V}_{u}|$. The inputs of the NN are $O_{0}\in\mathbb{R}^{|\mathcal{V}_0|}$, where $\mathcal{V}_{0}$ is the set of the inputs indexed from $1$ to $|\mathcal{V}_{0}|$. The output consists of a single neuron as $O_{|\mathcal{U}|}\in\mathbb{R}$. The NN layers' outputs are: 
\begin{equation}
    O_{u} = G_{u} (W_{u} O_{u-1} + b_{u}), \mspace{80.0mu} \forall u\in\mathcal{U},
\end{equation}
\noindent where $G_{u}\colon\mathbb{R}^{|\mathcal{V}_u|} \rightarrow \mathbb{R}^{|\mathcal{V}_u|}, u\in\mathcal{U}$, are the non-linear activation functions, and $W_{u}\in\mathbb{R}^{|\mathcal{V}_u|\times|\mathcal{V}_{u-1}|}$ and $b_{u}\in\mathbb{R}^{|\mathcal{V}_u|}$ are the weight and bias parameters of the NN layers.

\begin{figure}[t]
\vspace{-10pt}
    \centering
        \includegraphics[width=1.6in]{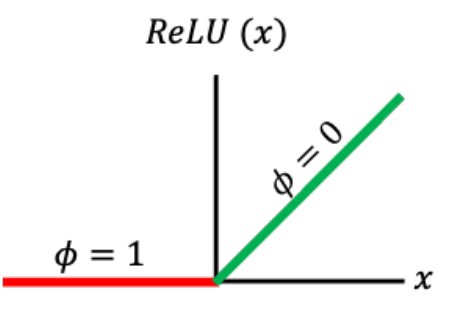}
	\caption[ReLU]{$ReLU(x) = \max\{0,x\}$ with binary variable $\phi\in \{0,1\}$, where $\phi = 1$ indicates the inactive red region and $\phi = 0$ the active green region. }
	\label{fig:ReLU_softsign}
 \vspace{-10pt}
\end{figure}

In~\cite{10012244}, we trained an NN to detect shared data manipulation using the $l_2$-norm of the primal residuals $||r^k||_2$ from the last 50 iterations as inputs. When the NN outputs $O^{|\mathcal{U}|} < 0$, the NN flags an attack. We chose $G_{u}(x) = ReLU(x) = \max\{0,x\}$ (see Fig.~\ref{fig:ReLU_softsign}) as the activation functions for the hidden layers and the identity for the output layer, i.e., $G_{|\mathcal{U}|}(x) = x$. 

Building on the work in~\cite{10012244}, we realized that using inputs to the NN based on the SC condition~\eqref{eq:condition2} evaluated for multiple iterations yields a better detection accuracy compared to the primal residuals. Specifically, the NN's input at iteration $s$ is:
\begin{equation} \label{eq:NN_inputs}
\begin{aligned}
& [O_{0}]_{v} = (x^{k}_a  - 2z^{k-1} + z^{k-2})^\mathsf{T} (x^{k}_a - x_a^{k-1}), \\ 
& \qquad\qquad\qquad\qquad\qquad\quad k = s - |\mathcal{V}_{0}| + v, \forall v \in \mathcal{V}_0.  
\end{aligned}
\end{equation}

To formulate this NN embedding, we use the method proposed in~\cite{cacciola2023structured} that models an NN with activation functions $ReLU(x) = \max\{0,x\}$ as MILP constraints. We introduce binary variables $\phi_{u}\in \{0,1\}^{|V_{u}|},\forall u \in \mathcal{U}$, to indicate the operation region of the $ReLU$ function as shown in Fig.~\ref{fig:ReLU_softsign}. We also introduce nonnegative real variables $I_{u}^{+}\in \mathbb{R}^{|V_{u}|}_{\geq0}$ and $I_{u}^{-}\in \mathbb{R}^{|\mathcal{V}_{u}|}_{\geq0},\forall u \in \mathcal{U}$, to model the inputs of the hidden layer $u$, with constraints that permit only one of a particular neuron's variables to take a nonzero value. For each layer, the inputs of the activation functions are $I_{u}^{+} - I_{u}^{-} = W_{u} \, O_{u-1} + b_{u}$ and the outputs are $O_{u} = ReLU(I_{u}^{+} - I_{u}^{-}) = I_{u}^{+}$.

Using this notation, we formulate the NN as MILP at iteration $k$ as follows. For each hidden layer $u \in \mathcal{U}\setminus \{|\mathcal{U}|\}$, we enforce the following constraints:
\begin{subequations} \label{eq:NN_embedding}
\begin{align}
&     I_{u}^{+} - I_{u}^{-} = W_{u} \, O_{u-1} + b_{u},  \\
&     O_{u} = I_{u}^{+}, \\
&     [\phi_{u}]_{v} = 1 \rightarrow [I_{u}^{+}]_{v} \leq 0, \qquad\qquad\qquad \forall v \in \mathcal{V}_u, \label{eq:NN_disjunctive_1} \\
&     [\phi_{u}]_{v} = 0 \rightarrow [I_{u}^{-}]_{v} \leq 0, \qquad\qquad\qquad \forall v \in \mathcal{V}_u, \label{eq:NN_disjunctive_2}\\
&     I_{u}^{+} \geq 0, \; I_{u}^{-} \geq 0.
\end{align}
\end{subequations}
Constraints~\eqref{eq:NN_disjunctive_1}--\eqref{eq:NN_disjunctive_2} are disjunctive inequalities that ensure only one of same entry of $I_{u}^{+}$ and $I_{u}^{-}$ has a nonzero value. These constraints can be reformulated as an MILP using Big-M or SOS1 methods. We define the inputs of the NN using~\eqref{eq:NN_inputs} and the outputs as:
\begin{subequations} \label{eq:NN_output}
\begin{align}
    & O_{|\mathcal{U}|} = W_{|\mathcal{U}|} O_{|\mathcal{U}|-1} + b_{|\mathcal{U}|},  \\
    & O_{|\mathcal{U}|} \geq \epsilon.\label{eq:NN_output_noattack}
\end{align}
\end{subequations}
Constraint~\eqref{eq:NN_output_noattack} ensures that the NN does not detect the outputs of the bilevel problem. A small positive number $\epsilon$ in~\eqref{eq:NN_output_noattack} accounts for the feasibility tolerance of the solver.

Solving the bilevel problem with an embedded NN is hard and does not scale well with large systems. Moreover, the complexity of the problem increases as the number of the hidden layers increases. Methods for efficiently embedding NNs in optimization problems has been an increasingly studied topic in recent years. Other NN embedding models may produce stronger MILP formulation than~\eqref{eq:NN_embedding}~\cite{anderson2020strong}. In future work, we plan to explore other NN embedding models with the bilevel optimization to scale to larger systems.

Embedding an NN to avoid detection with the simple and feedback attacks is more challenging than the bilevel optimization attack. The simple and feedback attacks compute the shared variables for a single iteration. Solving the projection problem~\eqref{eq:embd_feedback} with an embedded NN ensures the next iteration is undetected. However, the inputs of the NN detection model will be dominated by manipulated shared data after a few iterations. This makes finding the next iteration's value by solving~\eqref{eq:embd_feedback} with the NN embedding~\eqref{eq:NN_inputs}--\eqref{eq:NN_output} challenging because the attacker needs to consider many iterations after the next iteration to ensure finding feasible solutions. On the other hand, the bilevel attack avoids this problem because this attack ensures termination after two iterations. We will show in Section~\ref{sec6:results} that embedding an NN with the simple and feedback attacks leads to infeasible solutions.

\section{Adversarially Trained Neural Network} \label{sec5:adv_nn}

\begin{figure}[t]
\vspace{-20pt}
    \centering
        \includegraphics[width=2.75in]{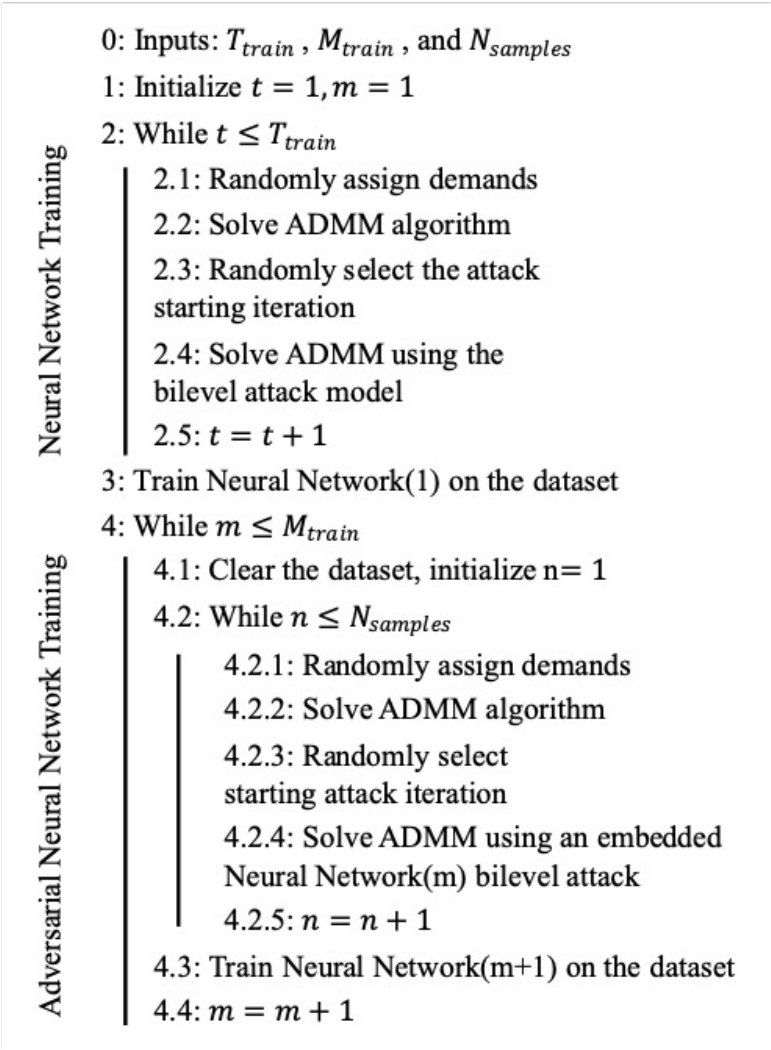}
         \vspace{-5pt}
	\caption[advflowchart]{Adversarial training framework flowchart consisting of two stages: (1) initial training in steps 2 and 3, and (2) adversarial training in step~4.}
	\label{fig:adv_flowchart}
 \vspace{-10pt}
\end{figure}

When augmented with~\eqref{eq:bilevel_additional}--\eqref{eq:NN_output}, the outputs of the bilevel optimization~\eqref{eq:bilevel} bypass both the detection condition~\eqref{eq:condition2} and the NN detection model presented in Section~\ref{subsec:embedding_NN}. To better detect such sophisticated data manipulation attacks, we develop an adversarial training framework. The framework uses the bilevel attack model to generate adversarial training samples as shown in Fig.~\ref{fig:adv_flowchart}. The framework's parameters are the number of initial training samples $T_{train}$, adversarial training iterations $M_{train}$, and samples for each adversarial training iteration $N_{samples}$.

Initially, we generate a dataset consisting of $T_{train}$ attacked and unattacked samples. We then train an NN to detect the attacked samples. The trained NN detects all the attacked samples even with a small number of layers and neurons. However, the bilevel attack with an embedded NN, as described in Section~\ref{subsec:embedding_NN}, can bypass this detection model. Accordingly, the second stage of the framework iteratively trains the NN on the output of the bilevel attack model. At each iteration, we generate $N_{samples}$ undetected attacked samples and retrain the NN on the new samples. The goal of this framework is to train the NN until the bilevel problem with an embedded NN becomes infeasible or too computationally expensive to solve, rendering this attack strategy ineffective. We focus on the bilevel attack model because the simple and feedback attacks fail to find adversarial samples.

\section{Simulation Results} \label{sec6:results}
This section presents the computational results of the data manipulation strategies and detection methods when solving OPF problems using the ADMM algorithm. We first show the impacts of the attack strategies on the solutions of the ADMM algorithm. Next, we demonstrate the effectiveness of the detection methods and compare their performance. We then present the results of embedding the detection methods into the data manipulation strategies. Finally, we show the results of the proposed adversarially trained NN framework in identifying data manipulation that embed the detection methods.

\subsection{Simulation Setup} 
We solved OPF problems with the case3\_lmbd, IEEE 14-bus, and IEEE 118-bus test cases from the PGLib-OPF library~\cite{PGLib}. We decomposed the systems into three areas and assumed that the attacker controls one of the areas. We set the tuning parameter $\rho = 100$ and terminate the ADMM algorithm when the $l_\infty$-norm of the primal residuals is less than $10^{-2}$.

Our implementation in Julia uses the PowerModelsADA library~\cite{alkhraijah2023powermodelsada} to solve OPF problems with the ADMM algorithm and the BilevelJuMP library~\cite{garcia2022bileveljump} to solve bilevel optimization problems. We trained NN models using the Flux library~\cite{Flux.jl-2018} and used Gurobi to solve all optimization problems. The simulations use an Intel Xeon CPU with 24 physical cores and 16GB memory. 

\subsection{Data Manipulation}
We first present the results of the data manipulation attacks. The attacker controls one of the areas and tries to steer the output of the ADMM to a suboptimal solution that increases the attacker’s local generation. The attacker uses three data manipulation strategies: (1) simple attack, (2) feedback attack, and (3) bilevel optimization attack. In both the simple and feedback attacks, the attacker finds the target value before starting the iterative process, whereas the bilevel attack finds the maximum achievable target values within the next two iterations after starting the attack.  

We used 100 runs to generate the results. In each run, we randomly select the load demands between 50\% and 150\% of the base demands. We also randomly select a starting iteration for the attack within the first hundred iterations. We use the \emph{optimality gap}, defined as the relative change in the objective function with and without data manipulation, to measure the impacts of the attacks. Table~\ref{tab:attack_result} shows the average optimal objective function value of the DCOPF problem and the average optimality gap when using each attack strategy. The three attacks find suboptimal solutions that increase the attacker's local generation. The optimality gap increases to 70\% for the 3-bus system and around 20\% for the 14- and 118-bus systems after the attacks.
\begin{table}[t]
    \centering
    \vspace{-20pt}
    \caption{The average optimal solution of the test cases over 100 runs and the optimality gap from the three attack strategies}
    \resizebox{\columnwidth}{!}{
    \begin{tabular}{|c||c||c|c|c|}\hline
        Case    &   Optimal Solution & Simple Attack & Feedback Attack & Bilevel Attack \\ \hline\hline
        3-bus   &      4388.2        &   70.2\%     &   70.2\%       &   70.2\%      \\\hline
        14-bus  &      7628.3        &   24.4\%     &   24.4\%       &   29.2\%      \\\hline
        118-bus &      125945.9      &   28.6\%     &   28.6\%       &   19.7\%      \\\hline
    \end{tabular}}
    \label{tab:attack_result}
\end{table}

\begin{table}[t]
\centering
\vspace{-8pt}
\caption{Accuracy of the detection methods}
\label{tab:detection_result}
\resizebox{\columnwidth}{!}{%
\begin{tabular}{|l||l||c|c|c|c|}
\hline
case & \begin{tabular}[c]{@{}l@{}}Detection\\ Method\end{tabular} & \begin{tabular}[c]{@{}c@{}}No \\ Attack\textsuperscript{1} \end{tabular} & \begin{tabular}[c]{@{}c@{}}Simple \\ Attack\textsuperscript{2} \end{tabular} & \begin{tabular}[c]{@{}c@{}}Feedback \\ Attack\textsuperscript{2}\end{tabular} & \begin{tabular}[c]{@{}c@{}}Bilevel \\ Attack\textsuperscript{2}\end{tabular} \\ \hline\hline
\multirow{2}{*}{3-bus} & DC & 100\% & 98.5\% & 100\% & 100\% \\ \cline{2-6} 
 & NN & 100\% & 100\% & 100\% & 100\% \\ \hline
\multirow{2}{*}{14-bus}  & DC & 100\% & 98.4\% & 100\% & 100\% \\ \cline{2-6} 
 & NN & 100\% & 100\% & 100\% & 100\% \\ \hline
\multirow{2}{*}{118-bus} & DC & 100\% & 98.5\% & 100\% & 94.2\% \\ \cline{2-6} 
 & NN & 97.8\% & 99.2\% & 99.9\% & 94.2\% \\ \hline
\end{tabular}}

\vspace{4pt}

{\textsuperscript{1}True negative. \textsuperscript{2}True positive.}
\vspace{-20pt}
\end{table}

\subsection{Detection Methods} 
We present the results of detection methods and a trained NN detection model. We checked the detection condition for all iterations until convergence. For the NN detection, our previous work~\cite{10012244} shows that we can achieve high detection accuracy by increasing the number of hidden layers. The work in~\cite{10012244} uses 8 hidden layers with 50 inputs corresponding to the mismatches of the shared variables over the last 50 iterations. Here, we show that we can maintain high detection accuracy with fewer hidden layers while considering fewer prior iterations as inputs to the NN.

As described in~\eqref{eq:NN_inputs}, we use a NN with 4 hidden layers and inputs from the last 10 iterations of the ADMM algorithm before convergence. We train the NN by generating 4000 samples of the three attack strategies and normal ADMM solutions. We then test the results using 1000 samples to produce the statistics shown in Table~\ref{tab:detection_result}.

The analytical detection condition avoids false negatives when the value of $\epsilon$ is appropriately selected. We set $\epsilon = 0.1$ in the detection condition for the 3- and 14-bus test systems and $\epsilon = 5$ for the 118-bus system to avoid false negative, i.e., flagging an attack for unattacked samples.

The analytical detection condition and the NN detection model successfully identify most of the attacks, with accuracy above 97\% for most of the test cases and often achieving above $99\%$ accuracy. Although the NN models use 10 iterations as inputs, their performance is close to the detection condition, which uses all the shared data from the first to the last iteration as inputs. Moreover, we can enhance the accuracy of the NN models by increasing the number of hidden layers and inputs. However, there is no control over the false negative results of the NN outputs, which can be seen in the 118-bus test system.

\begin{table}[t]
\centering
\vspace{-20pt}
\caption{Success rate of the attack strategies \\ with embedded detection methods}
\resizebox{\columnwidth}{!}{%
\begin{tabular}{|l||l||c|c|}
\hline
case & Detection Method & Feedback Attack & Bilevel Attack \\ \hline\hline
\multirow{3}{*}{3-bus} & DC & 100\% & 100\% \\ \cline{2-4} 
 & NN &  0\% & 72\% \\ \cline{2-4} 
 & NN+DC & 0\% & 44\%  \\ \hline
\multirow{3}{*}{14-bus} & DC &  100\% &  0\%   \\ \cline{2-4} 
 & NN &  0\%  & 0\% \\ \cline{2-4} 
 & NN+DC &  0\% &  0\% \\ \hline
\multirow{3}{*}{118-bus} & DC &  100\% &  0\% \\ \cline{2-4} 
 & NN & 0\% &  0\%  \\ \cline{2-4} 
 & NN+DC &  0\% &  0\%  \\ \hline
\end{tabular} \label{tab:embedding_result}
}
\vspace{-5pt}
\end{table}

\subsection{Embedding Detection Methods}
The high detection accuracy demonstrated in Table~\ref{tab:detection_result} can be compromised if the attacker knows the detection models. To show this, we use the two attack strategies with the embedded detection model described in Section~\ref{sec4:embed_detection}. We solve the models with 100 instances of the three test systems while varying the loads. We observe that the solver may take a very long time to find a solution due to the complexity of the problem. We set a maximum time limit of 100 seconds for the bilevel attack and 5 seconds for the feedback attack (since the attacker solves the bilevel problem once, but repeatedly solves projection problems for each iteration). 

The success rate of the two attacks is summarized in Table~\ref{tab:embedding_result}. The results indicate that knowing the detection models is not enough to ensure a successful attack. For the 14- and 118-bus systems, the attack strategies fail to find solutions in all instances when embedding the trained NN model. Moreover, the feedback attack always bypasses the two detection conditions, while the bilevel attack fails with the 14- and 118-bus systems. On the other hand, the feedback attack fails to bypass the NN detection model for the reason we previously discussed at the end of Section~\ref{sec4:embed_detection}.

\subsection{Adversarial Training}

To improve the NN detection accuracy, we use the adversarial training framework described in Section~\ref{sec5:adv_nn}. We use the 3-bus system to demonstrate the effectiveness of the framework because the attack models fail to avoid detection with the other two test systems. We iteratively trained the NN on the outputs of the bilevel attack models with an embedded NN. At each iteration, we solve the bilevel problem with $N_{samples} =100$ samples and collect the successful samples to retrain the NN. We use the bilevel attack model considering only the NN model and then using both the NN and the detection condition. 

\begin{figure}[t]
    \centering
    \vspace{-20pt}
        \includegraphics[width=3.3in]{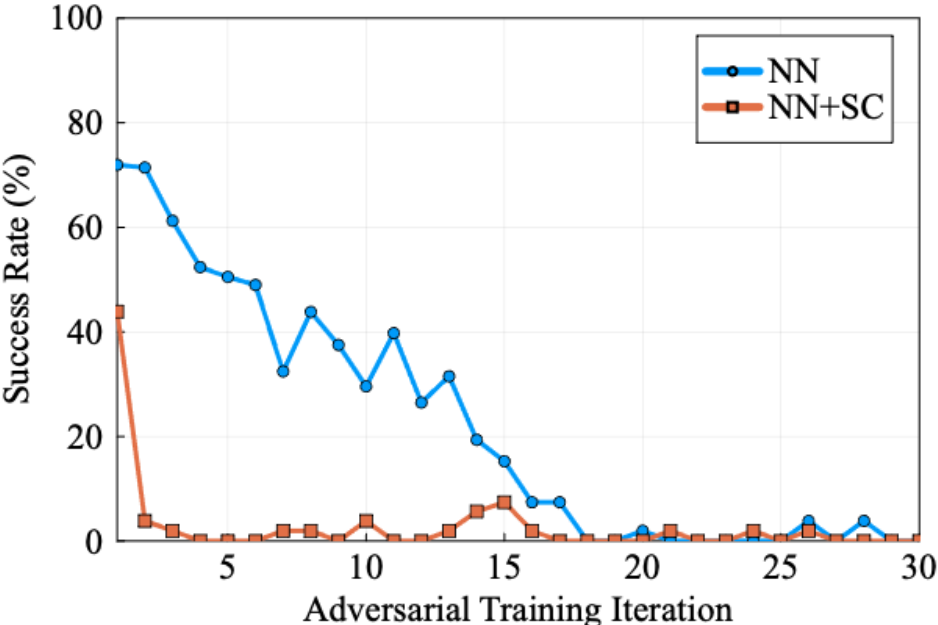}
	\caption[advresult]{Success rate of the bilevel attack with embedded detection methods over the adversarial training iterations. The blue line indicates the result of the bilevel attack with an embedded neural network (NN) and the red line with an embedded neural network and the detection condition (NN+DC).}
	\label{fig:adv_result}
    \vspace{-10pt}
\end{figure}

The results of the adversarial training are shown in Fig.~\ref{fig:adv_result}. The NN successfully detects almost all attacks (more than 99.9\%) with $M_{train} =16$ iterations. We also noticed that the detection accuracy improved more quickly when using the detection condition alongside the NN. Thus, the detection condition increases the efficiency of the training process by focusing on a subset of all the possible data manipulation scenarios to enhance the detection accuracy.

The results show that in a few instances, the bilevel attack finds solutions that are undetectable by the adversely trained NN. The framework does not guarantee that there is no possible bilevel attack solution. However, the possibility of finding a successful bilevel attack within 100 seconds is very low (i.e., less than 0.1\% with the setup in this paper).

\section{Conclusion} \label{sec7:conclusion}
Distributed algorithms have many advantages for coordinating systems with multiple agents. Future power systems with thousands or millions of controllers may coordinate their operations by solving OPF problems using distributed algorithms. However, as we demonstrate in this paper, distributed algorithms may be vulnerable to shared data manipulation attacks that drive the algorithm to suboptimal solutions. Since agents repeatedly solve the same subproblems, we exploit the subproblem structure to detect shared data manipulations by deriving a sufficient condition. We also show that a sophisticated attacker with knowledge of the detection methods may avoid detection by embedding the detection conditions in their attacks. We then use an adversarially trained NN framework to enhance the detectability of data manipulation strategies even when the attacker knows the detection methods. 

The proposed framework in this paper is based on embedding the detection methods into the attacker's data manipulation strategies. The embedding of the detection methods involves solving mixed-integer programs with non-convex quadratic constraints, which are computationally challenging. Accordingly, the results indicate that embedding the detection model into the attacker's problem does not scale well to large test systems. Our future work aims to develop embedding models that are more scalable through approximating or using stronger MILP models for the detection methods. Moreover, the attacker's problem is computationally difficult even without the NN embedding due to the need to ensure convergence of the distributed algorithm to the target solutions. Thus, our ongoing work aims to develop more scalable attack models that ensure convergence of the distributed algorithm. 
  
\section*{Acknowledgment}
The authors thank Scott Moura for insightful discussions on convergence theory for ADMM algorithms.

\appendix[Proof of the Sufficient Detection Condition] \label{sec:proof}

\begin{proof}
To derive the detection condition, let $\bar{f}(x) := f(x) + \delta_\mathcal{X}(x)$, the objective function of the local subproblem with the indicator function $\delta_\mathcal{X}(x)$, where $\delta_\mathcal{X}(x) = 0$ if $x\in\mathcal{X}$ and $\delta_\mathcal{X}(x) = \infty$ otherwise. Let $\bar{L}_\rho(x, z, y) := \bar{f}(x) + y^\mathsf{T}(A x + B z - c) + \frac{\rho}{2} ||A x + B z - c||^2_2$, the augmented Lagrange function~\eqref{eq:admm1} with the indicator function $\delta_\mathcal{X}(x)$. At iteration $k+1$, we have $x^{k+1} := \argmin~\bar{L}_\rho(x, z^{k}, y^{k})$. Since $f$ is lower semi-continuous over the set $\mathcal{X}$, and $\mathcal{X}$ satisfies the LICQ at $x^{k+1}$~\cite[Def. 12.4]{nocedal1999numerical}. Since $x^{k+1}$ minimizes $\bar{L}_\rho(x, z^k, y^k)$, then $0 \in \partial \bar{L}_\rho(x^{k+1}, z^{k}, y^{k})$, the subgradient of the objective function of the local subproblem evaluated at $x^{k+1}$~\cite[Theorem 8.15]{rockafellar2009variational}. Thus,
\begin{align}
& 0 & \in &~\partial \bar{L}_\rho(x^{k+1}, z^{k}, y^{k}), \nonumber \\
& & = &~\partial [\bar{f}(x^{k+1}) + (y^k)^\mathsf{T}(A x^{k+1} +Bz^{k} - c) \nonumber \\ &&& + \frac{\rho}{2} ||A x^{k+1} +Bz^{k} - c||_2^2], \nonumber \\
&  & = &~\partial \bar{f}(x^{k+1}) + A^\mathsf{T} (y^k) + \rho A^\mathsf{T}  (A x^{k+1} + B z^k - c), \nonumber \\
&  & = &~\partial \bar{f}(x^{k+1}) + A^\mathsf{T} (y^k) \nonumber \\ &&& + \rho A^\mathsf{T} \left[\frac{1}{\rho} (y^{k+1} - y^k) - B z^{k+1} + c+ B z^k - c\right], \nonumber \\
&  & = &~\partial \bar{f}(x^{k+1}) + A^\mathsf{T} (y^{k+1}) - \rho A^\mathsf{T} (B z^{k+1} - B z^k), \nonumber
\end{align}
\noindent where the fourth equality uses the dual update~\eqref{eq:admm3} to substitute for $Ax^{k+1} = \frac{1}{\rho} (y^{k+1} - y^{k}) - Bz^{k+1} +c$. This implies that $x^{k+1}$ also minimizes $\bar{f}(x) + (y^{k+1})^\mathsf{T} A x - \rho (B z^{k+1} - B z^k)^\mathsf{T} A x$. Thus, we have
\begin{align}
    & \bar{f}(x^{k+1}) + (y^{k+1})^\mathsf{T} A x^{k+1} - \rho (B z^{k+1} - B z^k)^\mathsf{T} A x^{k+1} \leq \nonumber \\ & \qquad \bar{f}(\tilde{x}) + (y^{k+1})^\mathsf{T} A \tilde{x} - \rho (B z^{k+1} - B z^k)^\mathsf{T} A \tilde{x}_i, \label{eq:condition1}
\end{align}
for any $\tilde{x}\in\mathbb{R}^n$. Replacing $k$ with $k-1$, we have
\begin{equation}
\begin{aligned}
    & \bar{f}(x^{k}) + (y^{k})^\mathsf{T} A x^{k} - \rho (B z^{k} - B z^{k-1})^\mathsf{T} A x^{k} \leq \\ & \quad\qquad \bar{f}(\tilde{x}) + (y^{k})^\mathsf{T} A \tilde{x} - \rho (B z^{k} - B z^{k-1})^\mathsf{T} A \tilde{x}, \end{aligned}\label{eq:condition1_k}
\end{equation}
for any $\tilde{x}\in\mathbb{R}^n$. Letting $\tilde{x} = x^k$ in~\eqref{eq:condition1}, $\tilde{x} = x^{k+1}$ in~\eqref{eq:condition1_k}, and combining both inequalities, we obtain the following:
\begin{align}
& \bar{f}(x^{k+1}) + (y^{k+1})^\mathsf{T} A x^{k+1} - \rho (B z^{k+1} - B z^k)^\mathsf{T} A x^{k+1}  \nonumber \\ 
& \qquad + \bar{f}(x^{k}) + (y^{k})^\mathsf{T} A x^{k} - \rho (B z^{k} - B z^{k-1})^\mathsf{T} A x^{k} \leq  \nonumber \\ 
& \qquad \bar{f}(x^{k}) + (y^{k+1})^\mathsf{T} A x^{k} - \rho (B z^{k+1} - B z^k)^\mathsf{T} A x^{k} + \nonumber \\ 
& \qquad + \bar{f}(x^{k+1}) + (y^{k})^\mathsf{T} A x^{k+1} - \rho (B z^{k} - B z^{k-1})^\mathsf{T} A x^{k+1}. \nonumber
\end{align}
The values of local objective functions with the local constraints $\bar{f}(x^k)$ and $\bar{f}(x^{k+1})$, which are private information, cancel out, and only the shared variables remain. Rearranging the terms and substituting for the dual variables, we obtain
\begin{align}
&\begin{aligned} & \!(y^{k+1})^\mathsf{T}\! (A x^{k+1} \!\!-\! A x^{k}) - \rho (B z^{k+1} \!\!-\! B z^k)^\mathsf{T} (A x^{k+1} \!\!-\! A x^{k})&  \nonumber  \\ 
& \qquad\qquad\quad\;\;\;\;\;\, - (y^{k})^\mathsf{T} (A x^{k+1} - A x^{k}) &\nonumber \\  
& \qquad\qquad\quad\;\;\;\;\;\, + \rho (B z^{k} - B z^{k-1})^\mathsf{T} (A x^{k+1} - A x^{k}) \leq 0,  \nonumber \end{aligned} \\
& (y^{k+1} - y^{k})^\mathsf{T} (A x^{k+1} - A x^{k}) \nonumber  \\ 
& \qquad - \rho (B z^{k+1} - 2 B z^k + B z^{k-1})^\mathsf{T} (A x^{k+1} - A x^{k}) \leq 0, \nonumber \\
& \rho (A x^{k+1} + B z^{k+1} - c )^\mathsf{T} (A x^{k+1} - A x^{k}) \nonumber \\ 
& \qquad - \rho (B z^{k+1} - 2 B z^k + B z^{k-1})^\mathsf{T} (A x^{k+1} - A x^{k}) \leq 0,  \nonumber \\
& (A x^{k+1} + 2 B z^k - B z^{k-1} - c )^\mathsf{T} (A x^{k+1} - A x^{k}) \leq 0.
\end{align}
\noindent We use the dual update~\eqref{eq:admm3} in the third inequality to substitute $y^{k+1} - y^{k} = \rho ( A x^{k+1} + B z^{k+1} - c)$. Factoring $A$ and substituting for  $\hat{z}^k = 2 B z^k - B z^{k-1} - c$ yields the detection condition. 
\end{proof}

\bibliographystyle{IEEEtran}
\bibliography{IEEEabrv,references.bib}

\begin{thebibliography}{10}
\providecommand{\url}[1]{#1}
\csname url@samestyle\endcsname
\providecommand{\newblock}{\relax}
\providecommand{\bibinfo}[2]{#2}
\providecommand{\BIBentrySTDinterwordspacing}{\spaceskip=0pt\relax}
\providecommand{\BIBentryALTinterwordstretchfactor}{4}
\providecommand{\BIBentryALTinterwordspacing}{\spaceskip=\fontdimen2\font plus
\BIBentryALTinterwordstretchfactor\fontdimen3\font minus \fontdimen4\font\relax}
\providecommand{\BIBforeignlanguage}[2]{{%
\expandafter\ifx\csname l@#1\endcsname\relax
\typeout{** WARNING: IEEEtran.bst: No hyphenation pattern has been}%
\typeout{** loaded for the language `#1'. Using the pattern for}%
\typeout{** the default language instead.}%
\else
\language=\csname l@#1\endcsname
\fi
#2}}
\providecommand{\BIBdecl}{\relax}
\BIBdecl

\bibitem{molzahn2017survey}
D.~K. Molzahn, F.~D{\"o}rfler, H.~Sandberg, S.~H. Low, S.~Chakrabarti, R.~Baldick, and J.~Lavaei, ``A survey of distributed optimization and control algorithms for electric power systems,'' \emph{IEEE Trans. Smart Grid}, vol.~8, no.~6, pp. 2941--2962, 2017.

\bibitem{ALKHRAIJAH2022108297}
M.~Alkhraijah, C.~Menendez, and D.~K. Molzahn, ``Assessing the impacts of nonideal communications on distributed optimal power flow algorithms,'' \emph{Electric Power Syst. Res.}, vol. 212, p. 108297, 2022, {\rm presented at} \emph{22nd Power Syst. Comput. Conf. (PSCC 2022)}.

\bibitem{7762902}
J.~Duan, W.~Zeng, and M.~Chow, ``Resilient distributed {DC} optimal power flow against data integrity attack,'' \emph{IEEE Trans. Smart Grid}, vol.~9, no.~4, pp. 3543--3552, 2018.

\bibitem{9539887}
Z.~Cheng and M.~Chow, ``Resilient collaborative distributed {AC} optimal power flow against false data injection attacks: {A} theoretical framework,'' \emph{IEEE Trans. Smart Grid}, vol.~13, no.~1, pp. 795--806, 2022.

\bibitem{9762779}
Y.~Yang, G.~Raman, J.~Peng, and Z.~Ye, ``Resilient consensus-based {AC} optimal power flow against data integrity attacks using {PLC},'' \emph{IEEE Trans. Smart Grid}, vol.~13, no.~5, pp. 3786--3797, 2022.

\bibitem{tpec2022}
R.~Harris, M.~Alkhraijah, D.~Huggins, and D.~K. Molzahn, ``On the impacts of different consistency constraint formulations for distributed optimal power flow,'' in \emph{Texas Power and Energy Conf. ({TPEC})}, 2022.

\bibitem{harris_molzahn-hicss2024}
R.~Harris and D.~K. Molzahn, ``Detecting and mitigating data integrity attacks on distributed algorithms for optimal power flow using machine learning,'' in \emph{57th Hawaii Int. Conf. Syst. Sci. (HICSS)}, 2024.

\bibitem{10012244}
M.~Alkhraijah, R.~Harris, S.~Litchfield, D.~Huggins, and D.~K. Molzahn, ``Analyzing malicious data injection attacks on distributed optimal power flow algorithms,'' in \emph{54th North American Power Symp. ({NAPS})}, 2022.

\bibitem{anomaly_survey}
G.~Pang, C.~Shen, L.~Cao, and A.~V.~D. Hengel, ``Deep learning for anomaly detection: A review,'' \emph{ACM Comput. Surv.}, vol.~54, no.~2, 2021.

\bibitem{munsing2018cybersecurity}
E.~Munsing and S.~Moura, ``Cybersecurity in distributed and fully-decentralized optimization: Distortions, noise injection, and {ADMM},'' \emph{\rm arXiv:1805.11194}, 2018.

\bibitem{molzahn_hiskens-fnt2019}
D.~K. Molzahn and I.~A. Hiskens, ``A survey of relaxations and approximations of the power flow equations,'' \emph{Found. Trends Electric Energy Syst.}, vol.~4, no. 1-2, pp. 1--221, 2019.

\bibitem{boyd2011distributed}
S.~Boyd, N.~Parikh, E.~Chu, B.~Peleato, and J.~Eckstein, ``Distributed optimization and statistical learning via the alternating direction method of multipliers,'' \emph{Found. Trends Mach. Learn.}, vol.~3, no.~1, 2010.

\bibitem{6917065}
A.~Kargarian, Y.~Fu, and Z.~Li, ``Distributed security-constrained unit commitment for large-scale power systems,'' \emph{IEEE Trans. Power Syst.}, vol.~30, no.~4, pp. 1925--1936, 2015.

\bibitem{rockafellar2009variational}
R.~T. Rockafellar and R.~J. Wets, \emph{Variational Analysis}.\hskip 1em plus 0.5em minus 0.4em\relax Springer Science \& Business Media, 2009, vol. 317.

\bibitem{nocedal1999numerical}
J.~Nocedal and S.~J. Wright, \emph{Numerical Optimization}.\hskip 1em plus 0.5em minus 0.4em\relax Springer, 1999.

\bibitem{8391733}
A.~Hauswirth, S.~Bolognani, G.~Hug, and F.~Dörfler, ``Generic existence of unique lagrange multipliers in {AC} optimal power flow,'' \emph{IEEE Control Syst. Lett.}, vol.~2, no.~4, pp. 791--796, 2018.

\bibitem{gopinath2022}
S.~Gopinath and H.~L. Hijazi, ``Benchmarking large-scale {ACOPF} solutions and optimality bounds,'' in \emph{IEEE Power \& Energy Society General Meeting (PESGM)}, 2022.

\bibitem{RepresentationAmet1981}
J.~Fortuny-Amat and B.~McCarl, ``A representation and economic interpretation of a two-level programming problem,'' \emph{J. Oper. Res. Soc.}, vol.~32, no.~9, pp. 783--792, 1981.

\bibitem{vielma2011modeling}
J.~P. Vielma and G.~L. Nemhauser, ``Modeling disjunctive constraints with a logarithmic number of binary variables and constraints,'' \emph{Math. Prog.}, vol. 128, pp. 49--72, 2011.

\bibitem{cacciola2023structured}
M.~Cacciola, A.~Frangioni, and A.~Lodi, ``Structured pruning of neural networks for constraints learning,'' \emph{\rm arXiv:2307.07457}, 2023.

\bibitem{anderson2020strong}
R.~Anderson, J.~Huchette, W.~Ma, C.~Tjandraatmadja, and J.~P. Vielma, ``Strong mixed-integer programming formulations for trained neural networks,'' \emph{Math. Prog.}, vol. 183, no. 1-2, pp. 3--39, 2020.

\bibitem{PGLib}
{IEEE PES Task Force on Benchmarks for Validation of Emerging Power System Algorithms}, ``The {P}ower {G}rid {L}ibrary for benchmarking {AC} optimal power flow algorithms,'' \emph{\rm arXiv:1908.02788}, Aug. 2019.

\bibitem{alkhraijah2023powermodelsada}
M.~Alkhraijah, R.~Harris, C.~Coffrin, and D.~K. Molzahn, ``\mbox{PowerModelsADA}: A framework for solving optimal power flow using distributed algorithms,'' \emph{IEEE Trans. Power Syst.}, vol.~39, no.~1, pp. 2357--2360, 2024.

\bibitem{garcia2022bileveljump}
J.~D. Garcia, G.~Bodin, and A.~Street, ``{BilevelJuMP.jl}: {M}odeling and solving bilevel optimization in {J}ulia,'' \emph{\rm arXiv:2205.02307}, 2022.

\bibitem{Flux.jl-2018}
M.~Innes \emph{et~al.}, ``Fashionable modelling with {Flux},'' \emph{\rm arXiv:1811.01457}, 2018.

\end{thebibliography}

\end{document}